# Mechanical modulation of the Gutenberg-Richter b-value in hierarchical fault networks

Bin Chen[1,2*], Mingze Zhao[1], Puyu Cao[3]

[1]Department of Engineering Mechanics, Zhejiang University, Hangzhou, China

[2]Key Laboratory of Soft Machines and Smart Devices of Zhejiang Province, Hangzhou, China

[3]Safety Key Laboratory of Sichuan Province, School of Mechanics and Aerospace Engineering, Southwest Jiaotong University, Chengdu, Sichuan 611756, China

* To whom correspondence should be addressed: chenb6@zju.edu.cn

## Abstract

The Gutenberg-Richter (G-R) law is seismology's most robust empirical regularity, yet a unified mechanical explanation for its slope—the $b$-value—remains a subject of ongoing debate. Here, we propose a framework suggesting that the $b$-value is mechanically modulated by the hierarchical architecture of fault networks. By coupling fracture mechanics with hierarchical network topology, we derive a generalized closed-form expression: $b = \frac{3}{3+2\gamma}\frac{\log_{10}k}{\log_{10}q}$. We further link these topological parameters to rate-and-state friction theory, illustrating that $k$ and $q$ can be mechanistically constrained by the competition between elastic stiffness, frictional healing, and energy dissipation. The empirical universality of $b\approx1$ may arise as the system balances these competing mechanisms, tending to reside near a mechanically balanced state ($k\approx q$). Furthermore, the dynamic decoupling of $k$ and $q$ under tectonic loading drives the system away from this critical state, providing a mechanical explanation for observed $b$-value anomalies. The stress-drop scaling exponent $\gamma$ enters as a modulated perturbation about the constant-stress ($\gamma$=0) zeroth-order baseline: it shifts $b$ only where source-scaling data independently constrain $\gamma$ away from zero. By establishing a mechanistic connection between macroscopic statistical variations and microscopic mechanical processes, this framework may provide a new perspective for interpreting fault dynamics.

## I. Introduction

The Gutenberg-Richter law, $log_{10}^{N}(\geq M) = a - bM$, introduced by Gutenberg and Richter in 1944 [1], is the quantitative backbone of statistical seismology. The persistent stability of its slope, $b\approx1$, across vastly divergent tectonic settings—from shallow strike-slip faults to deep subduction megathrusts [2-3]—implies a universal underlying physical principle. Over decades, two dominant interpretations have emerged. Scholz's rock-fracture experiments established the "stress meter" interpretation, where $b$ decreases with increasing differential stress [4-5]. In parallel, Aki linked $b$ to the fractal dimension $D_f$ of fault planes (b ≈ $D_f/2$), providing a geometric perspective [6]. Despite their empirical success, a unified, closed-form derivation that bridges fault-network mechanics with these macroscopic statistical laws has remained elusive.

Theoretically, self-organized criticality (SOC) has been the dominant paradigm for explaining power-law event-size statistics [7-8]. However, canonical SOC models, such as the Olami-Feder-Christensen (OFC) automaton, are purely numerical and lack analytical expressions for the $b$-value, often treating the fault as a homogeneous grid. More recent analytical efforts, such as those by Sornette et al. (2026), have successfully derived the G-R distribution from single-fault rupture physics by distinguishing self-arresting from runaway ruptures [9]. Yet, many of these frameworks primarily focus on the isolated dynamics of a single rupture, leaving the collective, hierarchical behavior of a branching fault network poorly understood.

Crucially, quite a few previous network models idealize faults as purely velocity-weakening surfaces. In reality, active fault zones are highly heterogeneous "mosaics" [10], composed of velocity-weakening asperities separated by velocity-strengthening creeping zones. Laboratory experiments and theoretical frameworks demonstrate that velocity-strengthening zones, often associated with fault gouge, accommodate significant aseismic slip and act as barriers that terminate rupture propagation, thereby dissipating strain energy [11-12]. Numerical simulations incorporating Rate-and-State Friction (RSF) [13-14] have further shown that velocity-strengthening barriers can arrest ruptures and partition strain energy between seismic and aseismic release [15]. This fundamental rheological heterogeneity challenges traditional assumptions of strict seismic-moment conservation and constant fracture energy across scales [16].

Here, we construct a finite hierarchical network model of a heterogeneous fault system. To capture the hierarchical cascade, we parameterize the system with three variables: the threshold branching ratio $k$, the area amplification factor $q$, and a stress-drop scaling exponent $\gamma$. Bridging the gap between continuum mechanics and network topology, our framework explicitly links the macroscopic $b$-value to microscopic mechanical processes. Crucially, we show that $k$ and $q$ are not static geometric constants, but are dynamically modulated by mechanical constraints. By coupling network topology with RSF stability theory, we derive a generalized analytical solution for the $b$-value. We show that the baseline $b\approx1$ reflects a mechanically balanced baseline state ($k \approx q$, $\gamma$=0) used as a zeroth-order reference, while departures from this state provide a mechanistic explanation for observed $b$-value anomalies.

## II. Theory

### A. Hierarchical network topology and the fault mosaic Model

We conceptualize the crustal fault system as a finite, self-similar hierarchical network comprising $N$ levels. Level 1 represents the smallest elementary micro-fractures, while Level $N$ represents the largest macroscopic ruptures. The network is governed by two fundamental geometric parameters: the threshold branching ratio $k$ and the area amplification factor $q$. A hierarchical network of 3 levels is illustrated in Figure 1.

Following the fault-mosaic model [10], we treat the junction between level $i-1$ and level $i$ as a velocity-strengthening creeping zone. The branching ratio, $k$, is defined mechanically as the number of level $i-1$ micro-ruptures to breach the creeping barrier. This barrier acts as an energy sink that dissipates a fraction of the incoming energy.The number of ruptures at level i, denoted as $n_{\mathrm{i}}$, scales as

$n_{\mathrm{i}} = k^{N-i}$. (1)

Taking the base-10 logarithm gives

$i = N - \frac{\log_{10}^{n_{\mathrm{i}}}}{\log_{10}^{k}}$. (2)

The rupture area at level $i$, denoted as $A_i$, expands self-similarly with an area amplification factor, denoted as $q$, given by

$A_{\mathrm{i}} = A_1 q^{i-1}$. (3)

It is important to note that $k$ and $q$ are not static geometric constants and are subject to a dual geometric-mechanical modulation. Geometrically, they can be governed by the width of the damage zone, the density of micro-fractures, and the three-dimensional connectivity of the fault network. Mechanically, however, the effective normal stress, pore fluid pressure, and the rheological state of the creeping zones can dynamically "shield" or "activate" these geometric nodes. Therefore, $k$ and $q$ are fundamentally effective topological parameters—representing the geometric network as "filtered" by mechanical conditions under a specific stress state.

### B. Scale-dependent fracture mechanics and derivation of the generalized G-R law

We anchor our theory in classical fracture mechanics [17]. For a crustal rupture of area $A$ and characteristic length $L$, we first adopt the idealized planar assumption such that $L \propto A^{1/2}$. A stress drop $\Delta\tau$ yields an average slip $D \propto \Delta\tau L/\mu$, with μ being the shear modulus. The resulting seismic moment $M_0$ consequently scales as

$$M_0 = \mu AD \propto \Delta\tau A^{3/2}. \quad (4)$$

This scaling is consistent with the standard definition of seismic moment and the Hanks-Kanamori moment-magnitude relation [18]. Note that the effects of fractal, non-planar rupture surfaces on the moment-area scaling can be readily absorbed into an effective area-filling coefficient, as discussed in Section III on geometric renormalization.

While broad source-scaling analyses demonstrate that the apparent stress remains roughly constant over a wide magnitude range [19], systematic deviations from this constant-stress-drop scaling can occur possibly due to scale-dependent fracture energy and fault roughness [20, 21], particularly over specific magnitude ranges or in highly heterogeneous fault zones. To account for these potential deviations, we parameterize a generalized scale-dependent stress drop, denoted as $\Delta\tau_i$, by introducing an exponent $\gamma$. such that

$$\Delta\tau_{\mathrm{i}} = \Delta\tau_1 q^{\gamma(i-1)}. \quad (5)$$

Substituting Eqs. (3) and (5) into Eq (4), the seismic moment at level *i*, denoted as $M_{0,i}$, becomes

$$M_{0,\mathrm{i}} \propto q^{(3/2+\gamma)(i-1)}. \quad (6)$$

Using the Hanks-Kanamori scale [18], the corresponding magnitude is given by

$$Mw_{,\mathrm{i}} = (1 + 2\gamma/3)(i-1)\log_{10}q + \mathrm{const}. \quad (7)$$

Substituting *i* from Eq. (2) into Eq. (7) and rearranging into the standard Gutenberg-Richter form, we obtain the generalized closed-form expression for the *b*-value, given by

$$b = 3/(3 + 2\gamma)\frac{\log_{10}k}{\log_{10}q}, \quad (8)$$

which reveals that the *b*-value is a metric jointly governed by network topology (*k*, q) and fracture energetics (γ). According to Eq. (8), *b*=1 when *k*=*q* with γ=0.

**C. Dynamic evolution of network parameters under tectonic loading**

The network parameters *k*, *q*, and *γ* in the model evolve dynamically in response to the tectonic loading process, exhibiting a fundamental bifurcation between two distinct mechanical phases (Figure 2). During the tectonic loading phase, progressive locking leads to strong elastic coupling across the fault and also suppresses aseismic slip, minimizing the damping effect of the creeping zones. Elastic coupling facilitates the synchronized pile-up of micro-ruptures, thereby decreasing $k$ ($dk/d\sigma<0$). Simultaneously, because strain energy cannot be released via distributed creep, it accumulates as undissipated elastic energy. When the fault finally fails, this energy drives a runaway rupture that expands across contiguous asperities, increasing $q$ ($dq/d\sigma>0$). For illustrative purposes, we then parameterize this evolution as $k = k_0 e^{-\alpha(\sigma-\sigma_0)}$, where $k_0$ is the critical number of precursor ruptures at $\sigma = \sigma_0$ and $\alpha$ is a constant greater than 0, and $q = q_0 e^{\beta(\sigma-\sigma_0)}$, where $q_0$ is the area amplification factor at $\sigma = \sigma_0$ and $\beta$ is a constant greater than 0. Physically, the exponential form of $k(\sigma)$ represents the reduction in the number of piled-up micro-ruptures as elastic coupling strengthens, whereas the exponential form of $q(\sigma)$ reflects the multiplicative buildup of undissipated elastic energy, as stronger coupling suppresses dissipation during each pile-up event. In the small-stress limit, such a formulation can reduce to the linear $b$-stress relation observed in laboratory acoustic emission experiments [5]. Substituting these into Eq. (8), the $b$-value will become a stress-dependent dynamic variable, given by

$$b = 3/(3 + 2\gamma)\left(\log_{10}^{k_0} - \alpha\sigma log_{10}e\right)/\left(\log_{10}^{q_0} + \beta\sigma log_{10}\mathrm{e}\right). \quad (9)$$

Under the assumption that $\gamma$ varies much more slowly than the exponential changes in $k$ and $q$, Eq. (9) implies a monotonic decrease of $b$ with increasing tectonic stress $\sigma$. This mathematically demonstrates that tectonic loading would drive the $b$-value downward (Figure 2).

Conversely, during the post-seismic relaxation phase, the abrupt stress drop triggers vigorous after-slip. The high creep rate acts as an energy sink, damping individual slip events. This ongoing dissipation forces slip fronts to densely pile up again, restoring a high $k$. Meanwhile, the extensive energy dissipated during this high-$k$ pile-up process leaves minimal residual energy to drive runaway expansion, suppressing $q$. This negative feedback may tend to keep the network statistically residing near $k \approx q$ (with $\gamma \approx 0$), which, via Eq. (8), sets $b \approx 1$ as a zeroth-order long-term mean rather than a locked value (Figure 2).

This temporal bifurcation can explain the spatial observations across different fault segments (Figure 3). For a given observational window, a lower shear strain rate implies that the region is predominantly locked and accumulating strain energy (loading-dominated). This corresponds to a low $k$ and high $q$ state, yielding a lower $b$-value with Eq. (8). In contrast, a higher strain rate implies a creeping-dominated region where energy is continuously dissipated through aseismic slip (relaxation-dominated). This corresponds to a high $k$ and low $q$ state, yielding a higher $b$-value with Eq. (8). As seen from Figure 3, our framework can interpret the observed positive correlation between shear strain rate and $b$-value across global fault systems. Interestingly, the macroscopically locked Alpine fault exhibits anomalously high $b$-values in Figure 3; this paradox can be resolved by the $k-q$ decoupling, where extreme fault fragmentation and rapid healing yield an exceptionally high $k$ that overrides its large $q$. This also suggests that $b$ is a dynamic state variable governed by topological competition, rather than a simple proxy for macroscopic locking.

## III. Discussion

With Eq. (8), the $b$-value is redefined not as a static statistical exponent, but as a coupled physical parameter balancing rupture threshold ($k$), spatial extent ($q$), and fracture energetics ($\gamma$). Because rupture area scales linearly while seismic moment scales as $A^{3/2}$, merging $k$ micro-patches yields a disproportionate seismic moment (and strain energy) surplus compared to individual events. Here, the creeping zones play a pivotal role: acting as mechanical buffers and energy dissipators, they absorb this surplus, breaking seismic-moment conservation and driving the system to statistically reside near a mechanically balanced state ($k \approx q$) with $b \approx 1$ as a zeroth-order baseline. At this baseline, the stress-drop exponent $\gamma$ serves as a mechanical perturbation to the network geometry. Empirical source-scaling data constrain $\gamma \approx 0$ as a zeroth-order approximation [19], meaning the $b$-value is primarily a mechanical readout of the fault network geometry ($\frac{\log_{10} k}{\log_{10} q}$). However, in regimes with significant mechanical departures (e.g., strongly locked asperities approaching failure), $\gamma$ deviates from zero alongside the decoupling of $k$ and $q$, providing a mechanistic explanation for observed $b$-value curvature and precursory drops.

In our framework, $b \approx 1$ may correspond to a profound physical state: a scale-invariant, uniform pile-up across all hierarchical levels. Geometrically, when $k \approx q$, the spatial density of piled-up slip fronts remains constant across scales, achieving optimal two-dimensional space-filling. Dynamically, this self-similar state can represent a balance between the energy dissipated during barrier breaching and the elastic energy released during rupture expansion. Mechanically, micro-ruptures act as stress pumps, but the velocity-strengthening barrier dissipates incoming energy via aseismic slip. When $k$ events cumulatively breach this barrier do the patches merge, transforming the temporal cascade length ($k$) into macroscopic spatial extent ($q$) and triggering the mainshock.

Deviations from criticality ($b$=1) into area-dominated ($b$<1) or threshold-dominated ($b$>1) regimes naturally produce runaway or fragmented ruptures, respectively. During the interseismic phase, progressive locking decreases $k$ and increases $q$, pushing the system away from criticality and driving precursory $b$-value drops. Following a mainshock, vigorous afterslip speeds up the creeping zones, dissipating a large amount of energy. This negative feedback tends to restore $k$ and suppress $q$, biasing the network back toward the $k \approx q$ baseline and keeping $b$ near 1 as a zeroth-order long-term mean rather than a locked value. This feedback loop, combined with the empirical baseline of constant apparent stress across scales ($\gamma \approx 0$) [19], provides a framework to interpret the decade-scale precursory $b$-value drops observed before M9 earthquakes [23].

Crucially, fault systems may not be always constrained to evolve $k$ and $q$ in strictly opposite directions, but exhibit diverse correlated trends. For instance, as a fault system matures and becomes universally more fragmented, both $k$ and $q$ may increase proportionally, preserving the critical state ($b \approx 1$). Conversely, transient fluid injection dramatically enhances fault healing and also energy dissipation, driving $k$ up while suppressing $q$, leading to anomalously high $b$-values. During localized locking prior to a major rupture, the lack of dissipation drives $k$ down and $q$ up, causing precursory $b$-value drops ($b<1$). This decoupling can fundamentally transform the $b$-value from a mere statistical artifact into a dynamic state variable that maps the full spectrum of earthquake statistics across the ($k$, $q$) phase space.

To map our topological parameters to fault mechanics, we focus on the mechanical side of this dual modulation to evaluate the branching ratio $k$ and the area amplification $q$. These parameters are driven by energy budgets derived from fracture mechanics and RSF stability theory. In a creeping zone, continuous aseismic slip acts as an energy sink. This ongoing dissipation prevents seamless rupture propagation and forces slip fronts to pile up densely, where the geometry of sliding fronts is a direct response to mechanical loading, as suggested by the mesoscopic adhesive friction study [24]. Because the high creep rate dampens the effective slip increment ($\Delta u_0$), the net energy contributed by each micro-rupture $\Delta E_0 \sim \Delta\tau_0 D_c$ is suppressed, where $\Delta\tau_0 \sim \mu\Delta u_0/L_0$ is the elastic stress perturbation from a single micro-rupture, with $L_0$ being the characteristic length of a micro-rupture and $D_c$ is the critical slip distance. Consider that, to overcome the macroscopic fracture energy $\Gamma$, a large number of piled-up fronts accumulate so that we have

$$k \sim \frac{\Gamma}{\Delta E_0} \sim \frac{(a-b)_c \sigma_n \ln\left(V_{dyn}/V_{creep}\right)}{\mu \Delta u_0/L_0}, \quad (10)$$

where $(a-b)_c>0$ is the velocity-strengthening parameter of the creeping zone, $\sigma_n$ the effective normal stress, $V_{creep}$ the background aseismic slip rate, and $V_{dyn}$ the dynamic slip rate (with $V_{\mathrm{dyn}} \gg V_{\mathrm{creep}}$) by assuming the ratio of critical slip distance of creeping zones to that of microruptures is fixed across scales. In Eq. (10), a higher creep rate of a creep zone would increase $V_{dyn}$ and reduce $\Delta u_0$, which would drive $k$ to a larger value.

On the other hand, the potential for runaway propagation is governed by the competition between the elastic unloading stiffness of the surrounding medium ($K_{el}$ $\sim\mu/L_1$ with $L_1$ being the macroscopic rupture length), and the frictional weakening stiffness of the parent asperity ($K_{RSF}\sim(b-a)_{asp}\sigma_n/D_c$), where $(b-a)_{asp}>0$ characterizes the velocity-weakening nature of the parent asperity. Due to the energy dissipation of creep zones, we also introduce an energy retention factor, Λ, into the classical fracture criterion, given by

$$q \sim \Lambda(K_{RSF}/K_{el})^2 \sim \Lambda\left(\frac{(b-a)_{asp}\sigma_n/D_c}{\mu/L_1}\right)^2, \quad (11)$$

which suggests a small Λ due to the extensive energy dissipation of creeping zones would suppress $q$.

While our macroscopic strain-rate analysis captures the regional contrast in $b$-values (Figure 3), source-scaling studies illuminate the underlying mechanical architecture, the San Andreas Fault provides a compelling natural laboratory for validating this mapping. Crucially, a stark spatial heterogeneity is observed: Dreger et al. (2007) [25] documented that Mw 2.1 repeating earthquakes at Parkfield exhibit remarkably low spatial average stress drops (~2.5–5.6 MPa) but extremely high localized peaks (66.7–93.9 MPa), while Kim et al. (2016) [26] reported even higher localized peaks (69.5–94.7 MPa) alongside a time-invariant average of ~12 MPa. This massive contrast—extreme localized stress drops against a low regional average—is the direct observational signature of a fault populated by small, mechanically strong asperities embedded in a weak, creeping matrix [10]. It supports the premise of our fault-mosaic network: the macroscopic $b$-value is not dictated by a homogeneous stress state, but by the dynamic competition between high-threshold asperities and dissipative creeping zones.

Beyond spatial statistics, introducing a time axis allows our hierarchical network to function as a spatiotemporal branching process (Please see Supplemental Information). Here, we define $k_{aft}$ as the effective branching ratio for aftershocks, while $\lambda_{fore}>1$ and $\lambda_{aft}>1$ represent the characteristic temporal scaling factors between successive cascade generations for foreshocks and aftershocks, respectively. For foreshocks (bottom-up coalescence), the branching process is supercritical ($1<k$). This yields the inverse-Omori law [27, 28] with an acceleration exponent $p_{fore}=1-\log_{\lambda_{fore}}k<1$. Following the mainshock, the sudden stress drop triggers vigorous aseismic afterslip. For aftershocks (top-down triggering), the effective branching ratio is subcritical ($k_{aft}$

≤1), which yields the Omori-Utsu decay [29-32] with a relaxation exponent $p_{aft}=1-\log_{\lambda_{aft}} k_{aft} \geq 1$. Crucially, because these temporal exponents depend solely on the topological cascade ratios ($k$, $k_{aft}$) and the temporal scaling ($\lambda$), they are independent of the stress-drop scaling parameter $\gamma$. This demonstrates a physical orthogonality: the time-dependent cascade dynamics and the energy partitioning operate independently. Since $p_{aft} \geq 1$ and $p_{fore}<1$, it follows within our framework that $p_{fore}- p_{aft} <0$. This provides an analytical realization of the empirical observation that the inverse-Omori exponent for foreshocks is systematically smaller than the Omori exponent for aftershocks, consistent with regional waveform analyses in Japan [32] and with cascade-based explanations of foreshock acceleration [28]. We suspect that this topological asymmetry may serve as a diagnostic to separate purely cascade-driven sequences from those requiring transient aseismic or fluid-driven loading [33].

Although our mean-field network predicts orthogonality between the temporal exponent and the static $b$-value (as they depend on distinct topological parameters), real catalogs exhibit heterogeneous correlations [34]. We suggest that both can be modulated by a common underlying variable: the stress state of creeping zones [2, 5]. Tectonic loading drives $b$ downward while altering the cascade rate $\lambda$. Conversely, fluid-induced seismicity relaxes effective normal stress, elevating $b$ and increasing $\lambda$ [33]. Thus, the creeping zone is not a passive buffer, but an active stress-sensitive element whose dynamics determine whether $b$ and the temporal exponent appear coupled, decoupled, or anti-correlated in a given catalog.

Our framework may also provide a common physical basis for other classical empirical scaling laws. First, our framework provides a natural parameterization for Båth's law (the largest aftershock is typically ~1.2 magnitudes smaller than the mainshock [34]). Under the constant-stress approximation ($\gamma \approx 0$), Eq. (7) yields a direct relationship between the magnitude gap and the area amplification factor: $\Delta M \approx \log_{10} q$. If $q$ can be independently measured (e.g., from source-scaling relations or geodetic moment budgets) to be $\approx 16$, then Eq. (7) predicts $\Delta \mathrm{M} \approx 1.2$, consistent with Båth's law. Assuming the fault network operates near the mechanically balanced baseline, this implies a branching ratio of $k \approx 16$, suggesting that approximately 16 sub-ruptures are required to trigger a parent event. Furthermore, the upward curvature of the G-R distribution requires large-scale non-self-similarity. As ruptures approach strongly locked asperities, decreasing $k$ and increasing effective $\gamma$ lower the local $b$-value at the high-magnitude tail, producing an excess of large events. This matches observations in high-stress asperities [36]. However, we do not explicitly derive the empirical spatial scaling of aftershock zones established by Utsu [37], as mapping discrete hierarchies to continuous space requires incorporating complex Coulomb stress decay beyond our mean-field model.

It is important to distinguish the mean-field result in Eq. (8) from its geometrically renormalized counterpart. Real faults are non-planar, so rupture area need not scale as $L^2$. To leading order, geometric constraints can be absorbed into a connectivity factor $\chi_p \leq 1$ and an area-filling efficiency $\eta_A \leq 1$. Using effective parameters $k_{\mathrm{eff}}=\chi_p k$ and $q_{\mathrm{eff}}=\eta_A q$, and writing $L \propto A^{1/D_f}$ ($D_f \leq 2$ being the fractal dimension of rupture surface), the generalized exponent, dentoed as $b_{\mathrm{geo}}$, becomes $b_{geo} = \frac{3}{2(1+1/D_f+\gamma)} \frac{log_{10}k_{eff}}{log_{10}q_{eff}}$. In the ideal limit ($D_f=2$ and $\chi_p=\eta_A=1$), this reduces exactly to Eq. (8). This renormalization may help bridge our framework with first-principles rupture models (10, 15). While these numerical models capture local RSF dynamics, our topology may help bridge the gap to a macroscopic analytical framework.

Finally, our framework provides testable pathways to physically constrain $k$ and $q$ across different spatial scales. At the mesoscopic level, discrete numerical simulations and laboratory acoustic emission experiments offer direct, high-resolution measurements of $k$ and $q$. For macroscopic natural faults—where observations are inherently challenging—$k$ can be statistically inferred from high-quality, declustered seismicity catalogs, and $q$ can be approximated by jointly assessing geodetic locking distributions and finite-fault rupture areas. Although field quantification remains an ongoing challenge, explicitly defining these topological parameters here provides a concrete target for future observations, transitioning the mechanical modulation of the $b$-value from a theoretical framework into a falsifiable physical hypothesis.

## IV. Conclusion

We have derived the Gutenberg-Richter law analytically from a spatially heterogeneous, fractal hierarchical fault network. By treating velocity-strengthening creeping zones as mechanical buffers and energy dissipators, and coupling network topology with fracture mechanics, we obtained a generalized closed-form expression for the $b$-value. Our results demonstrate that tectonic loading drives the dynamic decoupling of $k$ and $q$, explaining the precursory $b$-value drops before critical failure. This suggests that the long-term stability of $b \approx 1$ reflects a mechanical equilibrium between elastic loading and creep dissipation ($k \approx q$). Transient mechanical deviations from this equilibrium provide a mechanistic basis for the observed b-value anomalies and G-R curvature.

Furthermore, by grounding temporal cascades in Rate-and-State Friction (RSF) scaling, we reveal that foreshock acceleration and aftershock decay are dual manifestations of pre-seismic progressive breaching and post-seismic afterslip. These temporal exponents are decoupled from static fracture energy, demonstrating an orthogonality between time-dependent friction and static energy release.

Finally, real fault networks are bidirectionally coupled; our framework can be generalized by making $k, q, \gamma$, and $\lambda$ level-dependent, direction-dependent and dynamically updated [10, 33]. Operationally, our topological framework transforms the $b$-value from a static statistical descriptor into a dynamic state variable. For densely recorded fault systems, future studies may diagnose local variations in $q$ via the dominant magnitude gap (e.g., the mainshock-largest aftershock difference) and map the $(k,q)$ trajectory in real time; however, this magnitude gap serves as a diagnostic indicator and is not used as an input to constrain the framework. This approach may offer a physics-based diagnostic tool to distinguish between creeping relaxation, fluid-induced swarms, and critically locked patches approaching failure. As the system approaches macroscopic failure and enters a highly nonlinear, critically accelerated state, bridging this macroscopic framework with real-time data assimilation—and recognizing the handover to LEFM at the nucleation patch [38]—can define the next frontier in operational forecasting. In parallel, integrating poroelastic deformation-diffusion coupling [39] can also be essential for capturing stress transients in fluid-sensitive environments and completing this physics-based forecasting pipeline.

**Acknowledgments**

This work was supported by the National Natural Science Foundation of China (Grant No.: 12372318).

**Author contributions:** B.C. designed research, B.C. performed research, B.C., M.Z. and P.C. analyzed data, B.C. wrote the manuscript.

**Competing Interest Statement:** The authors declare no competing interests.

# Figures

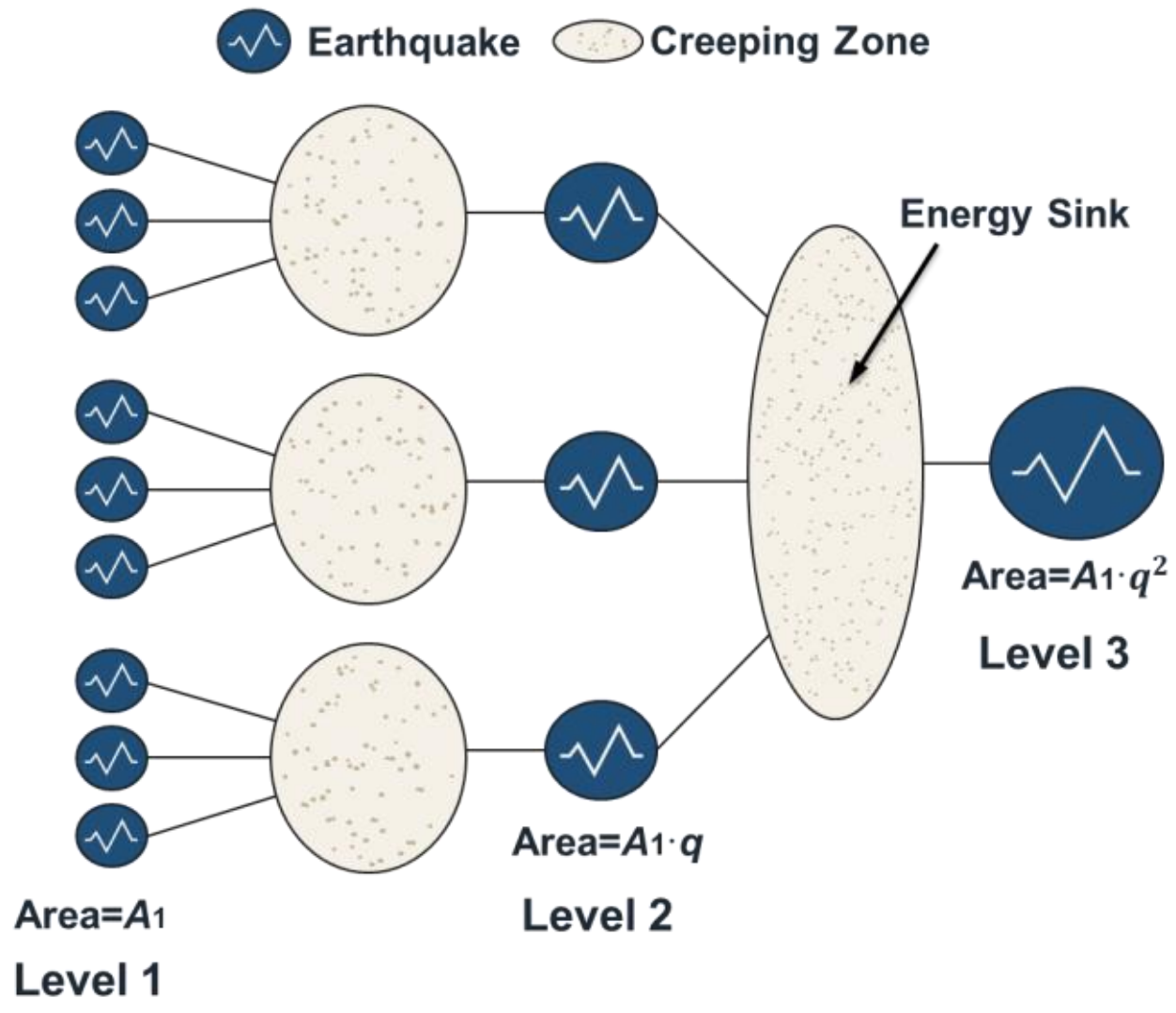

**Figure 1 Schematic of the hierarchical fault network and topological cascading with *N*=3**. The fault system is idealized as a discrete hierarchical network where micro-ruptures are separated by velocity-strengthening creeping zones. The creeping zones act as mechanical buffers and energy sinks, dissipating incoming elastic stress via aseismic slip and forcing slip fronts to densely pile up. A critical number of $k$ successive micro-ruptures must accumulate to overcome the breaching threshold. Once breached, the isolated patches coalesce into a larger parent asperity, amplifying the rupture area by a factor of $q$. This self-similar cascade repeats at progressively larger scales.

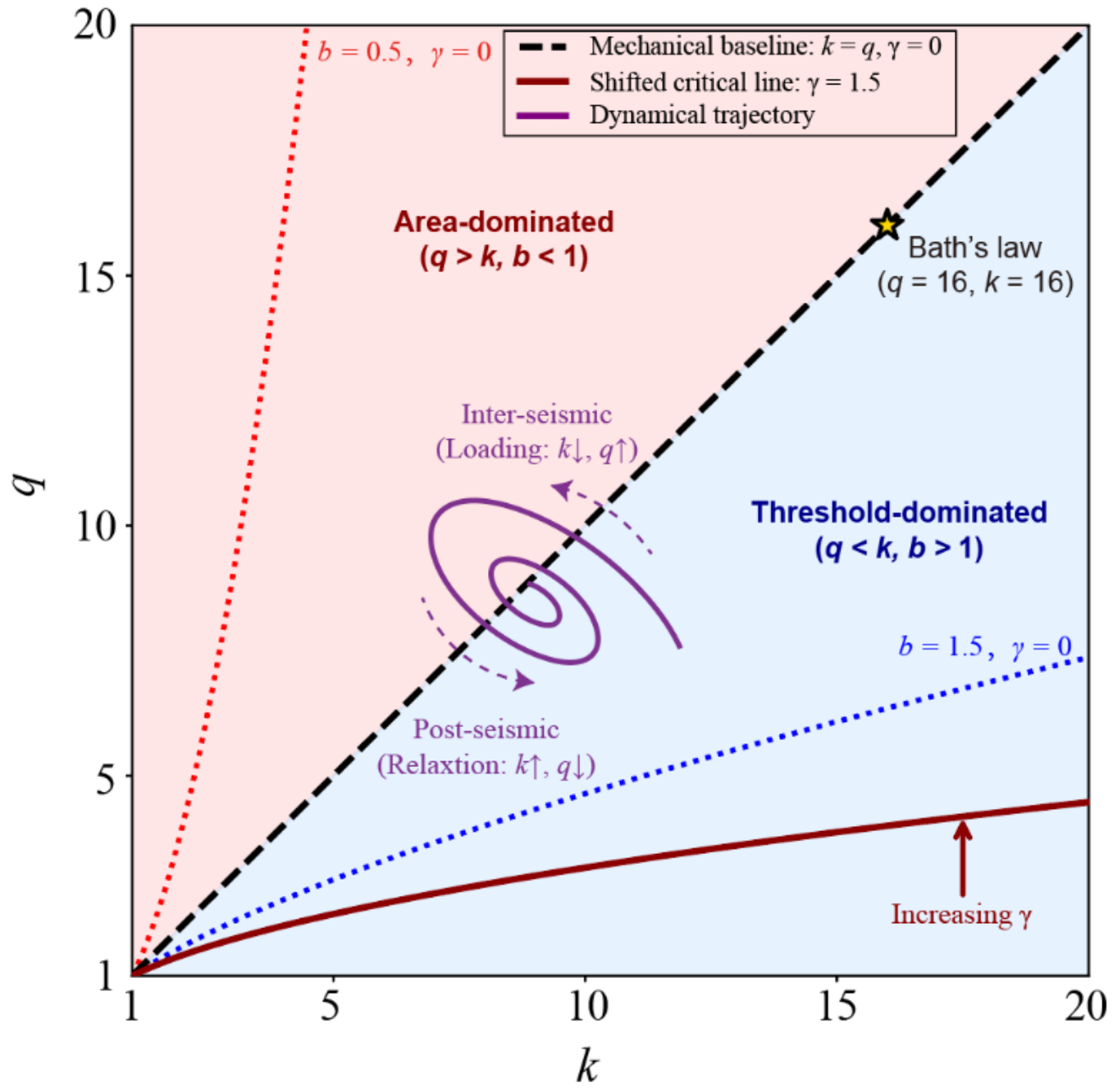

**Figure 2 Dynamical phase diagram in the (*k*, *q*) space illustrating the seismic cycle.** The black dashed line denotes the mechanically balanced baseline ($k$=$q$, yielding $b$=1 for $\gamma$=0). The background shading separates the area-dominated regime ($q$>$k$, yielding $b$<1) from the threshold-dominated regime ($k$>$q$, yielding $b$>1). The red solid line is an illustrative sensitivity curve showing how the baseline would shift if $\gamma$>0; it is not fitted to data. The gold star marks a candidate equilibrium ($k$≈$q$); the value $q$≈16 shown there is illustrative and is consistent with Båth's law ($\Delta M$≈1.2). The purple trajectory illustrates the earthquake cycle: tectonic loading drives the system to the upper-left (decreasing $k$, increasing $q$, and driving precursory $b$-value drops), while post-seismic afterslip tends to restore $k$ and suppress $q$, biasing the network back toward this baseline. The dashed contours ($b$=0.5,1.5) are drawn for $\gamma$=0, the zeroth-order case adopted throughout the analysis.

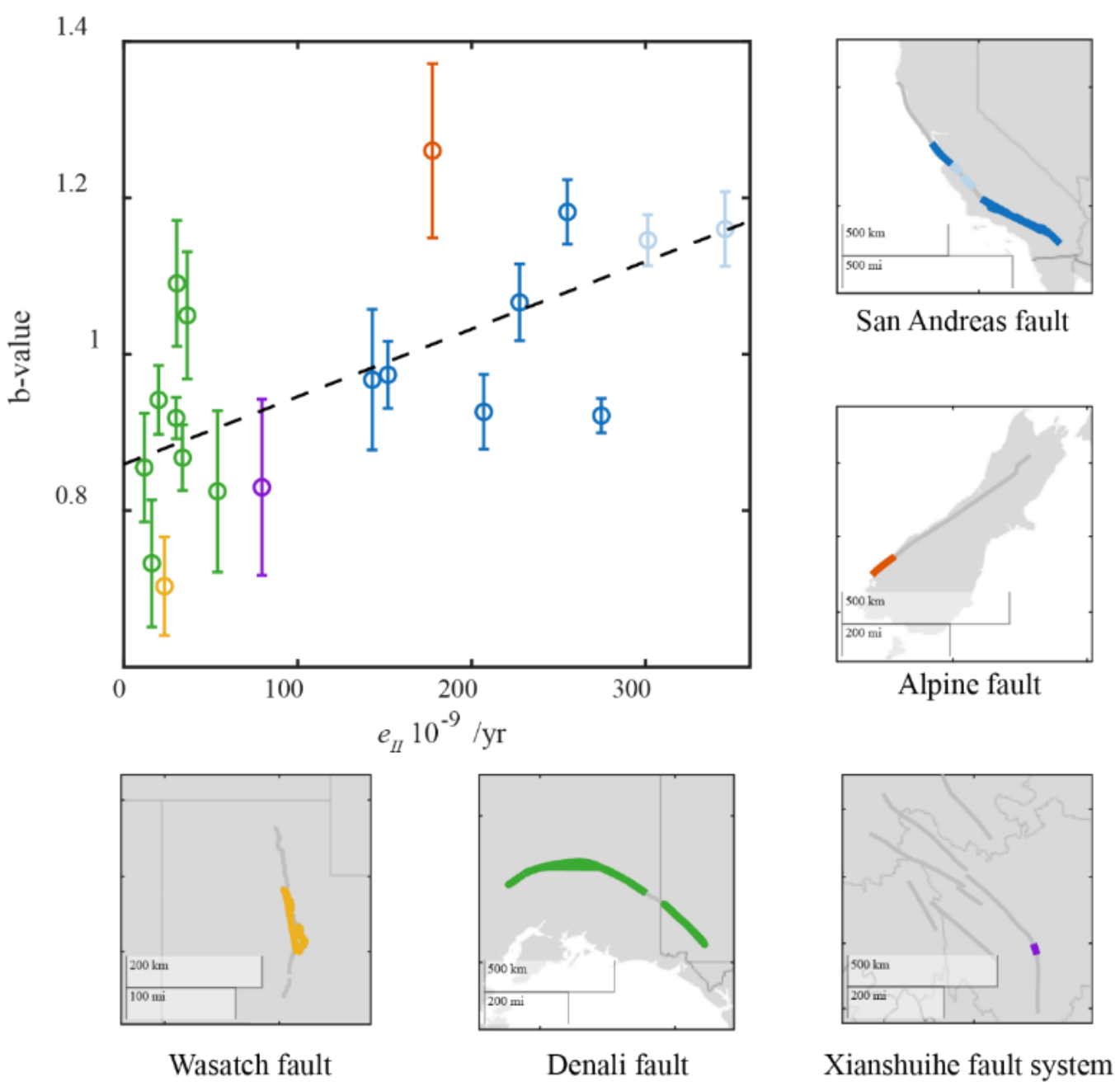

**Figure 3 Observed positive correlation between *b*-value and shear strain rate across global continental fault systems.** The scatter plot shows local *b*-values versus the second invariant of the shear strain rate, $e_{II}$, for several major continental strike-slip and normal faults. The black dashed line represents the linear regression, demonstrating a clear positive correlation. Specifically, the San Andreas fault hosts a well-documented creeping section, which is shown with light blue symbols, while its locked sections are shown with dark blue symbols. The Alpine, Wasatch, Denali, and Xianshuihe faults show no documented shallow creep and are entirely represented by dark-colored symbols. Map insets show the exact data extraction windows and scale bars. Earthquake catalogs for all five zones are retrieved from the USGS ComCat (ANSS) for 2005–2025 via the FDSN event web service. Crustal strain rates are taken from the GSRM v2.1 geodetic strain-rate model [22] on its global 0.1° grid. The detailed data analysis is provided in Supplemental Information.

# Supplementary Information

## S1. Temporal cascade dynamics

The temporal evolution of seismicity is modeled as a spatiotemporal branching process governed by Rate-and-State Friction (RSF) in the velocity-strengthening creeping zones.

### S1.1 Foreshock acceleration (Bottom-up convergent cascade)

During the pre-seismic phase, tectonic loading increases the local stress and the creep rate in the creeping zones, causing the time delay between generations, denoted as $\Delta t_m$, to shrink geometrically. We define $\Delta t_m = \tau_0 \lambda_{fore}^{-m}$ with $\tau_0$ being a time constant and $\lambda_{fore}>1$. Assuming the cascade spans a sufficiently large number of generations ($N \gg$ m), the time remaining until the mainshock at $t_f$ is given by

$$t_f - t_m \approx \tau_0 \frac{\lambda_{fore}^{-m}}{1-\lambda_{fore}^{-1}} \text{, (S1)}$$

leading to

$$\text{m}=- \log_{\lambda_{fore}} \frac{t_f - t_m}{\tau_0/(1-\lambda_{fore}^{-1})}. \text{ (S2)}$$

The number of events in the $m^{th}$ generation scales as $N_m \propto k^{-m}$. Substituting Eq. (S2) into the single-generation rate given by $R_m = \frac{N_m}{\Delta t_m} \propto (\frac{\lambda_{fore}}{k})^m$, we then obtain the inverse-Omori power law, given by

$$R(t) \propto (t_f - t)^{-\left(1-\log_{\lambda_{fore}} k\right)}, \text{ (S3)}$$

yielding the acceleration exponent, $p_{fore} = 1 - \log_{\lambda_{fore}} k$.

### S1.2 Aftershock relaxation (Top-down divergent cascade)

Following the mainshock, the sudden stress drop reduces the background creep rate, causing the time delays between successive cascading generations to stretch geometrically. In the late-time asymptotic regime ($m \gg 1$), we define the time interval between $(m-1)^{\text{th}}$ and $m^{\text{th}}$ generations as $\Delta t_m \propto \lambda_{aft}^m$, with $\lambda_{aft}>1$ is the temporal scaling factor. The time of the $m$-th generation aftershock is then given as

$$t_m - t_f = \tau_{0f}\, \lambda_{aft}^m, \text{ (S4)}$$

where $\tau_{0f}$ is the characteristic time scale of the first generation. The number of events in the $m^{th}$ generation scales as $N_m \propto k_{aft}^m$, with $k_{aft} \leq 1$ representing the average triggering ratio. The single-generation seismicity rate is then given by

$$R_m = \frac{N_m}{\Delta t_m} \propto (\frac{k_{aft}}{\lambda_{aft}})^m. \text{ (S5)}$$

Solving Eq. (S4) for the generation index $m$, we obtain:

$$m = \log_{\lambda_{aft}}^{\frac{t_m - t_f}{\tau_{0f}}}. \text{ (S6)}$$

Substituting Eq. (S6) into Eq. (S5) yields the Omori-Utsu decay law, given by

$$R(t) \propto (t - t_f)^{-\left(1 - \log_{\lambda_{aft}}^{k_{aft}}\right)}, \text{ (S7)}$$

leading to the relaxation exponent, $p_{aft} = 1 - \log_{\lambda_{aft}}^{k_{aft}}$.

**S2. Cascade scaling and Bath's law**

Bath's law states that the largest aftershock is typically 1.2 magnitudes smaller than the mainshock [34]. Consider that the mainshock occurs at level $N$, and the largest aftershock occurs at level $N$−1. Using the generalized magnitude scaling Eq. (7), the magnitude difference is given by

$$\Delta \mathrm{M} = M_{w,N} - M_{w,N-1} = (1 + (2/3)\gamma)\log_{10} q. \text{ (S8)}$$

For $q$ taken as an illustrative value of $\approx 16$ pending independent measurement (e.g., from source-scaling relations or geodetic moment budgets), Eq. (S8) predicts $\Delta \mathrm{M} \approx$ 1.2 in the $\gamma = 0$ limit, consistent with Båth's law.

**S3. Data analysis for the relation between *b* and crustal strain rate**

We estimate the relation between *b*-value and crustal strain rate for five major fault systems: the San Andreas fault, the Denali fault (Alaska), the Alpine fault (New Zealand), the Wasatch fault zone (Utah), and the Xianshuihe fault system (Sichuan, China). Fault traces are taken from the GEM Global Active Faults database [S1], where the San Andreas, Alpine, Wasatch, and Denali faults are selected by name and the Xianshuihe fault system is selected by corridor box selection (28.5–34.5°N/96–103°E) with a sinistral slip-type filter. Of these, the San Andreas fault hosts a well-documented creeping section [S2] that is represented in the Global Active Faults database [S1], which are used to classify its bins as creeping or locked; the Alpine, Wasatch, and Denali faults show no documented shallow creep and are treated as locked. Earthquake catalogs for all five zones are retrieved from the USGS ComCat (ANSS) for 2005–2025 via the FDSN event web service. Crustal strain rates are taken from the GSRM v2.1 geodetic strain-rate model [22] on its global 0.1° grid.

In the analysis, each fault system is partitioned into bins of 100 km length along the strike [S3], and earthquakes within 50 km of the fault trace are employed and assigned to bins according to their projected along-strike coordinate. The magnitude of completeness $M_c$ is estimated for each fault zone from the whole corridor catalog by a Kolmogorov–Smirnov goodness-of-fit test, and *b*-values are estimated within each bin by maximum likelihood [S4] using events with M⩾Mc. For statistical reliability, bins with fewer than 100 events above Mc are discarded; while given their sparser catalogs, the threshold is relaxed to 60 for Denali fault and 40 for the Alpine fault and Xianshuihe zones. The strain rate is taken as the median value of the strain rate within the bins, and the second invariant of the strain rate tensor, $e_{II}$, is employed, which is given by $e_{II} = \sqrt{e_1^2 + e_2^2}$, where $e_1$ and $e_2$ are two principal strain rates.

## Supplemental References